\documentclass[12pt,preprint]{aastex}

\shorttitle{The Spectrum of Earthshine}
\shortauthors{Woolf et al.}

\begin{document}


\title{The Spectrum of Earthshine: \\
   A Pale Blue Dot Observed from the Ground}


\author{N. J. Woolf and P. S. Smith}
\affil{Steward Observatory, Tucson AZ 85721}
\email{nwoolf@as.arizona.edu, psmith@as.arizona.edu}

\and

\author{W. A. Traub and K. W. Jucks}
\affil{Harvard-Smithsonian Center for Astrophysics, Cambridge MA 02138}
\email{wtraub@cfa.harvard.edu, kjucks@cfa.harvard.edu}



\begin{abstract}
We report the visible reflection spectrum of the integrated Earth, 
illuminated as it would be seen as an spatially-unresolved extrasolar 
planet.  The spectrum was derived from observation of lunar earthshine 
in the range 4800 to 9200\AA\  at a spectral resolution of about 600.  
We observe absorption features of ozone, molecular oxygen and water. 
We see enhanced reflectivity at short wavelengths from Rayleigh 
scattering, and apparently negligible contributions from aerosol and 
ocean water scattering.   We also see enhanced reflectivity at long 
wavelengths starting at about 7300\AA, corresponding to the well-known 
red reflectivity edge of vegetation due to its chlorophyll content; 
however this signal is not conclusive, due to the breakdown of our 
simple model at wavelengths beyond 7900\AA.
\end{abstract}




\section{Introduction}


Earthshine is sunlight which has reflected from the Earth onto the dark side of 
the moon and back again to Earth.  Here ``dark side'' refers to that portion of the
lunar surface which, at any instant, faces the Earth but does not face the sun.
Both earthshine, from the dark side of the moon, and moonlight, from the bright 
side of the moon, are transmitted through the same
air mass just prior to detection,  and thus suffer the same extinction and imposed 
absorption features.  Their ratio is the reflection coefficient of the Earth 
multiplied by a geometric factor, and by the ratio of phase effects of the 
moon. (We ignore weak additive Raman-shifted spectra from both bodies, i.e., 
the Ring effect.)  We report here the results of our effort to obtain a 
spectrum of the Earth as it would be seen from outside the solar system.  
Our goal is to pave the way for interpreting extrasolar terrestrial planet 
spectra which might be obtained in the future, for example as part of the 
Terrestrial Planet Finder (TPF) program.  

\section{Observations}

We observed the spectrum of earthshine and moonlight soon after sunset on 
June 24/25 2001 (25/26 UT) from Kitt Peak, Arizona, using the Steward 
Observatory 2.3m  telescope with its Boller and Chivens spectrograph, a CCD 
detector and a 300 line/mm grating.  The raw spectrum is shown in Fig.~1.
A spectral range from 4800 to 9200\AA\   
was recorded, at a resolution of about 600, or a spectral width of about 8-9\AA. 
 The telescope drive 
was set to lunar rate and was guided in right ascension, keeping the limb in 
the same position on the slit to within about a minute of arc in 
declination between the different observations.   At the time of these
observations, about 1/4 of the moon appeared sunlit, and 3/4 was nominally dark. 

The 1.5 arcsec wide spectrograph slit has a length of about 200 arcsec.  
It was set perpendicular 
to the lunar limb, with approximately half of its length on the limb and half 
on adjacent sky.  Both the sunlit and Earth-lit limb spectra were 
corrected for scattered light in the 
telescope by subtracting the adjacent sky spectrum; this amounted to about 2/3 
of the total signal in the earthshine spectrum.  Each earthshine spectrum was 
bracketed by moonlight spectra.

Both nights were partly cloudy, and the observations were made during breaks 
in the clouds.  Data from the first night have more noise than data from the 
second night, but are in qualitative agreement. The second night produced two 
independent sets of observations; the first was taken under more 
clear conditions, and is the one reported here.

The measured reflectivity of the integrated Earth is shown in Fig.~2.  
The reflectivity is the ratio of the earthshine to moonlight spectra, both 
sky-subtracted. We have also corrected for the color effect of lunar phase 
\citep{lane73}, which makes the apparent reflectivity about 16\% too 
strong in the blue;  this color bias was removed by multiplying the apparent 
reflectivity by a 8-segment piecewise-linear factor which ranged from 0.835 at 
5000\AA\   to 1.00 at 9000\AA.  

The reflectivity in Fig. 2 has also been corrected for CCD fringing.
Dome flats were not adequate to exactly match the lunar-spectrum fringing 
in the thinned
CCD chip because flexure in the instrument and/or telescope changed the
position of the fringes slightly.  Use of a quartz lamp internal to the 
spectrograph ensured that fringes were at the same location on the 
chip as for the lunar observations, but their amplitude was suppressed
because the lamp does not illuminate the slit with an f/9 beam as does
light that passes through the telescope.  We used the internal lamp 
to partially correct the fringes, by dividing by the lamp spectrum, 
but a noticeable fringe pattern remained.  Five
residual interference fringe packets in the red part of the spectrum were
least-squares fitted with  smooth gaussian envelope sinusoids and subtracted
from the data.  The fitting procedure removed most of the remaining
fringing, but there remains a 
residual noise which is larger in this region than elsewhere in the spectrum.  
The zero-mean subtracted component is explicitly displayed (offset in 
intensity) beneath the data in Fig.~2.

\section{Spectrum Model and Analysis}

Our model of the Earth's reflection spectrum is a simple box model in which 
each box represents a potentially major contributor to the final  spectrum.  
By using a box model, we ignore the interactions between components 
(e.g., reflections between clouds and ground, multiple scattering, etc.);
in other words, we focus on the net resulting spectrum, not the interplay 
of generating mechanisms.
The model has 9 independent parameters, each of which scales an independent 
physical component.  The standard-atmosphere component is discussed in 
\citet{traub02}, and the other components are discussed briefly here.

(1) Optical depth of standard atmospheric species H$_2$O, O$_2$, and O$_3$.  
The optical depth is calculated at very high spectral resolution, using a 
model atmosphere with thin layers from 0 to 100 km altitude, standard mixing 
ratio profiles of species, and Doppler and pressure broadening.  The resulting 
spectrum is smeared and sampled to match the experimental wavelength grid and 
resolution.

(2) Optical depth of stratospheric O$_3$.  Here O$_3$ is the only absorber, and  
atmospheric layers are limited to stratospheric altitudes, otherwise similar 
to the full standard atmosphere component.

(3) Rayleigh scattering.  This is a source term with 
$\lambda^{-4}$ functional form.

(4) Aerosol scattering.  This is also a source term but with 
$\lambda^{-1.3}$ form.

(5) Reflectivity of high clouds.  This is a constant reflectivity term, 
independent of wavelength, as from a very high cloud.

(6) Reflectivity of ocean water.  This is light which has penetrated the 
surface and has been scattered out of the water.  It has a 
$10^{-(\lambda - \lambda_0)/\Delta\lambda}$ 
form, giving the ocean its blue color, where $\lambda_0 = 5500$\AA\  
and $\Delta \lambda = 1300$\AA\   here.

(7) Reflectivity of ocean pigment.  This is ocean pigment due to chlorophyll 
and its products, here represented by a gaussian centered at 5500\AA\   and with 
full width at half maximum 650\AA, giving waters with phytoplankton a green 
color.  The longer-wavelength chlorophyll feature (item 8) is not seen in practice;
water absorbs quite strongly at these wavelengths (item 6).

(8) Reflectivity of vegetated land.  This is the reflectivity of vegetation 
containing chlorophyll, on land, dominated by a sharp rise in reflectivity 
for wavelengths longer than about 7200\AA, plus some smaller bumps at shorter 
and longer wavelengths.  It is modeled here by functional forms  approximating 
these features.   (If our eyes were sensitive to wavelengths longer than 7200\AA,
all vegetation would appear to be infrared bright, not green!)

(9) Reflectivity of low clouds and surface, below bulk of atmosphere.  
The nominally featureless reflectivity of land, low clouds, and the 2\% Fresnel 
reflectivity of the air-ocean interface are lumped into this single term, 
independent of wavelength.

In terms of our model, the two dominant components in the spectrum are the 
``high cloud'' continuum and the ``clear atmosphere'' spectrum of water, ozone, 
and oxygen.  All components are displayed in Fig.~2, where the linear sum of 
the 7 component spectra is equal to the smooth model function superposed on 
the data.  The main features of water, ozone, and oxygen are labeled; weak 
bands of water are not labeled.  The relative strength of each spectral 
component is proportional to the intrinsic reflectivity times the
effective projected area on Earth of that component.  
For example, suppose that the clear 
atmosphere spectrum arises entirely over water, which has a Fresnel 
reflectivity of 0.02, and that the continuum spectrum is from clouds, with a 
nominal reflectivity of 0.30.  Then the fact that we observe these components 
to have similar strengths indicates that the clear atmosphere area is roughly 
15 times greater than the cloudy area, i.e., the model Earth has a relatively 
clear sky.

Rayleigh scattered light from molecules is the next largest component.  
In our box model this component is multiplied by the transmission of the clear 
atmosphere, which is how it is displayed in Fig.~2.
The remaining components are relatively small, and carry less significance 
than the dominant three.  

The vegetation ``red edge'' signal in Fig.~2 is the
well-known signature of chlorophyll in land vegetation, described above.  
The earthshine 
spectrum shows increased reflectivity in the 7200 to 7900\AA\   region, suggestive 
of a vegetation signal in this range.  However the data do not show continued 
high reflectivity at longer wavelengths (7900 to 9200\AA), as would be expected for a 
vegetation reflection profile, making the interpretation ambiguous.  
Nevertheless, on the basis of the abrupt reflectivity increase at 7200\AA\   we 
suspect that the vegetation signal is indeed present, and that either the data 
or the model fail at longer wavelengths, masking this signal.

The three remaining components are weakest.  The ocean clear-water signal, 
the aerosol signal, and the ocean phytoplankton signal all make relatively 
small contributions to the spectrum.  Note that there is a superficial 
similarity between each of the blue continuum spectra (Rayleigh, ocean, and 
aerosol), however their wavelength variations are quite different when viewed 
over a wide range of wavelengths, and this allows the components to be 
independently determined, with relatively small correlation.

In summary, the earthshine spectrum shows strong evidence for cloud, ozone, 
water, oxygen, Rayleigh scattering, and very probably vegetation signals.  
Apparently weak components are aerosols, ocean water, and phytoplankton 
signals.

\section{Discussion}

In order to characterize an extrasolar terrestrial planet using visible light,
there are three main avenues of approach.  (1) The diurnal 
photometric variability could be measured. \citet{ford01} have shown 
that for an Earth-like planet, variation on the order of 100\% may be 
expected.  (2) The spectral reflectivity of the atmosphere could be measured. 
As shown here and in \citet{traub02}, absorption features due to 
ozone, water vapor, and molecular oxygen could be seen, as well as enhanced 
reflectivity due to Rayleigh and aerosol scattering.  (3) The spectral 
reflectivity of the surface could be measured.  As we have shown in Fig.~2. 
This potentially includes the reflectivity of clear ocean water (blue), 
phytoplankton in the ocean (green), and land vegetation (red).

The vegetation signal in Fig.~2 is relatively small, about 6\% above the 
nearby continuum.  But at the time of observation only about 17\% of the 
projected area was land, as shown in the simulated cloud-free Earth 
image\footnote{Earth simulation images are
available at http://www.fourmilab.ch/earthview/vplanet.html}
in Fig.~3.
Of this fraction of land,  only part is cloud-free and vegetated, 
so in principle the vegetation signal could be at least 6 times larger than 
we observe.   An infrared view of the whole 
earth\footnote{http://www.ssec.wisc.edu/data/comp/ir/} 
showed little or no cloud cover 
over the land masses illustrated in Fig.~3, so it is reasonable to expect 
that we might be seeing a large part of the available vegetation signal 
in our spectrum.

For example, this expectation is borne out in data taken by the GOME 
satellite\footnote{For a sample spectrum, see http://earth.esa.int/symposia/data/popp} 
in which the spectral region from about 3100 to 8000\AA\  was observed 
in a nadir-pointing mode, and the field of view was relatively 
small, and filled with vegetated land in northern Italy.  In that reflection 
spectrum, the vegetation signal is about 2.4 times larger than the adjacent 
continuum, showing that this is a potentially very large differential signal 
on a similarly vegetated extrasolar planet. 
 
The residual spectral mismatch  in Fig.~2 
in the range 7300-7900\AA\   suggests that yet more vegetation signal could 
be accommodated in the model, but that this may be thwarted by a bias 
in either the data or the model in the 7900 to 9200\AA\   region.
Although a more detailed reflectivity model of the composite Earth has yet to
be developed, we believe that the enhancement of reflectivity redward of
7300\AA\  is significant, and is an indicator of vegetation.

\section{Comments}

Our spectrum of the Earth as it would appear to an extrasolar observer
will be useful for learning how to analyze the spectra of extrasolar planets, 
discovering the key spectral features of Earth analogs, and determining the 
necessary spectral resolution and signal-to-noise ratios.  This spectrum illustrates 
both atmospheric and surface reflectivity features.  It was not feasible to 
look for variability because of the limited observing period.

We note that the reflectivity of the Earth has not been static throughout the 
past 4.5 Gyr.  In particular, oxygen and ozone become abundant perhaps 
only 2.3 Gyr ago, affecting the atmospheric absorption component of the reflection 
spectrum.  Then perhaps about 2.0 Gyr ago a green phytoplankton signal 
developed in the oceans.  About 0.44 Gyr ago, an extensive land plant 
cover developed, generating the red chlorophyll edge in the reflection 
spectrum \citep{lunine99}.  

An observer in a nearby stellar system would have been
able to use the oxygen and ozone absorption features to suspect the presence of life
on Earth anytime during the past 50\% of the age of the solar system.  The extrasolar
observer would have been able to use the chlorophyll red-edge reflection
feature to confirm the presence of life on Earth anytime during the most recent 10\%
of the age of the solar system.

As the Earth evolved there have also 
been significant changes in the temperature structure, the abundances of H$_2$O,
CO$_2$, CH$_4$, and perhaps the character of cloud coverage.  Some of these changes
may have been sudden and even oscillatory, as opposed to gradual.  Thus the spectrum
of the evolving Earth has shown dramatic changes (e.g., \citet{traub02}), 
and we may expect that similar signatures will be found on extrasolar terrestrial 
planets.

We suggest that further observations of Earthshine should be made.
One purpose is to validate our ability to characterize extrasolar 
terrestrial planets using visible wavelengths    In particular, it would 
be valuable to make earthshine observations from space as well as ground, to 
remove the complicating factor of observing through the Earth's atmosphere.  
It would also be valuable to observe different types of land and ocean and 
cloud cover on Earth, and furthermore to do so for a full diurnal cycle in 
order to test the spectroscopy and the prediction of variable photometry.







\acknowledgments

We thank our colleagues on the TPF biomarkers team study \citep{desmarais02} 
initiated by Charles Beichman and led by David Des Marais.   
We thank Felix Kogan, James Marr, Sara Seager, Jim Kasting, and an 
anonymous referee for advice. 
This work was supported in part by funds from Jet Propulsion Laboratory's 
TPF studies, to NW via the Lockheed-Martin corporation, 
and to WT and KJ via the Ball Aerospace corporation.




\clearpage


\begin{figure}
\figurenum{1}
\plotone{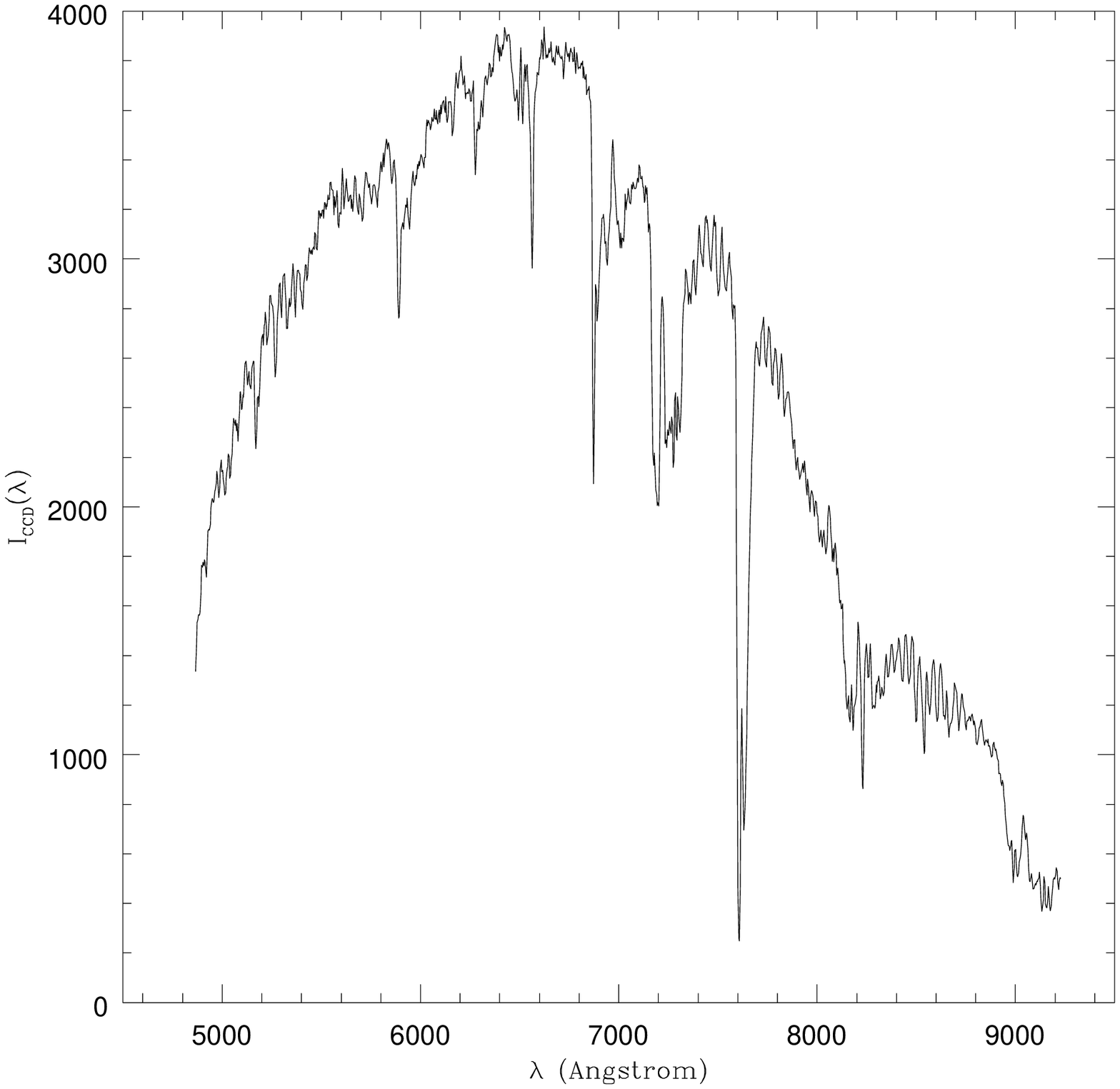}
\caption{Spectrum of earthshine, after sky subtraction.  
Ordinate is CCD data units. 
 A mix of solar features such as Na D line and H$\alpha$ 
is seen  along with terrestrial oxygen A and B bands, and many water vapor 
features.  CCD fringes disturb the spectrum from about 7200-9200\AA. 
}
\end{figure}

\clearpage 

\begin{figure}
\figurenum{2}
\plotone{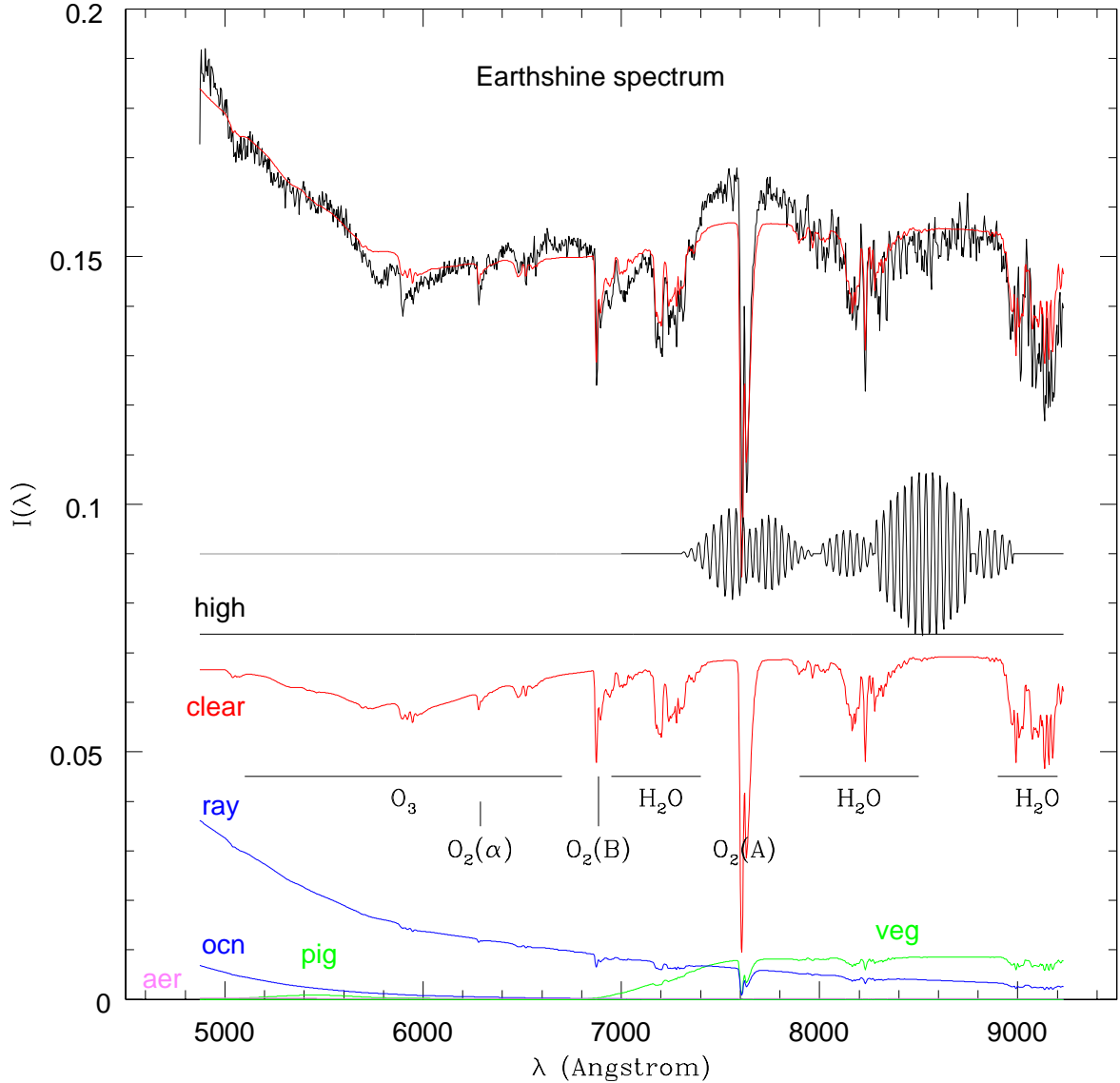}
\caption{The observed reflectivity spectrum of the integrated Earth, 
as determined from earthshine, is shown at the top, with a model spectrum 
superposed.  The reflectivity scale is arbitrary.   Five interference 
fringe packets (inset, offset but otherwise to scale) simulating CCD 
fringing were subtracted from the data.  Seven component spectra (below) 
were fitted and summed to produce the model spectrum. 
{\bf High} is reflectivity from a high cloud.   
{\bf Clear} is the clear-atmosphere transmission.
{\bf Ray} is  Rayleigh scattered light.  
{\bf Veg} is the vegetation reflection spectrum from land chlorophyll plants. 
{\bf Ocn} is the blue spectrum from subsurface ocean water.  
{\bf Aer} is aerosol scattered light, here negligible. 
{\bf  Pig} is the green pigmented phytoplankton reflection of ocean waters, 
also negligible here. 
}
\end{figure}

\clearpage 

\begin{figure}
\figurenum{3}
\plotone{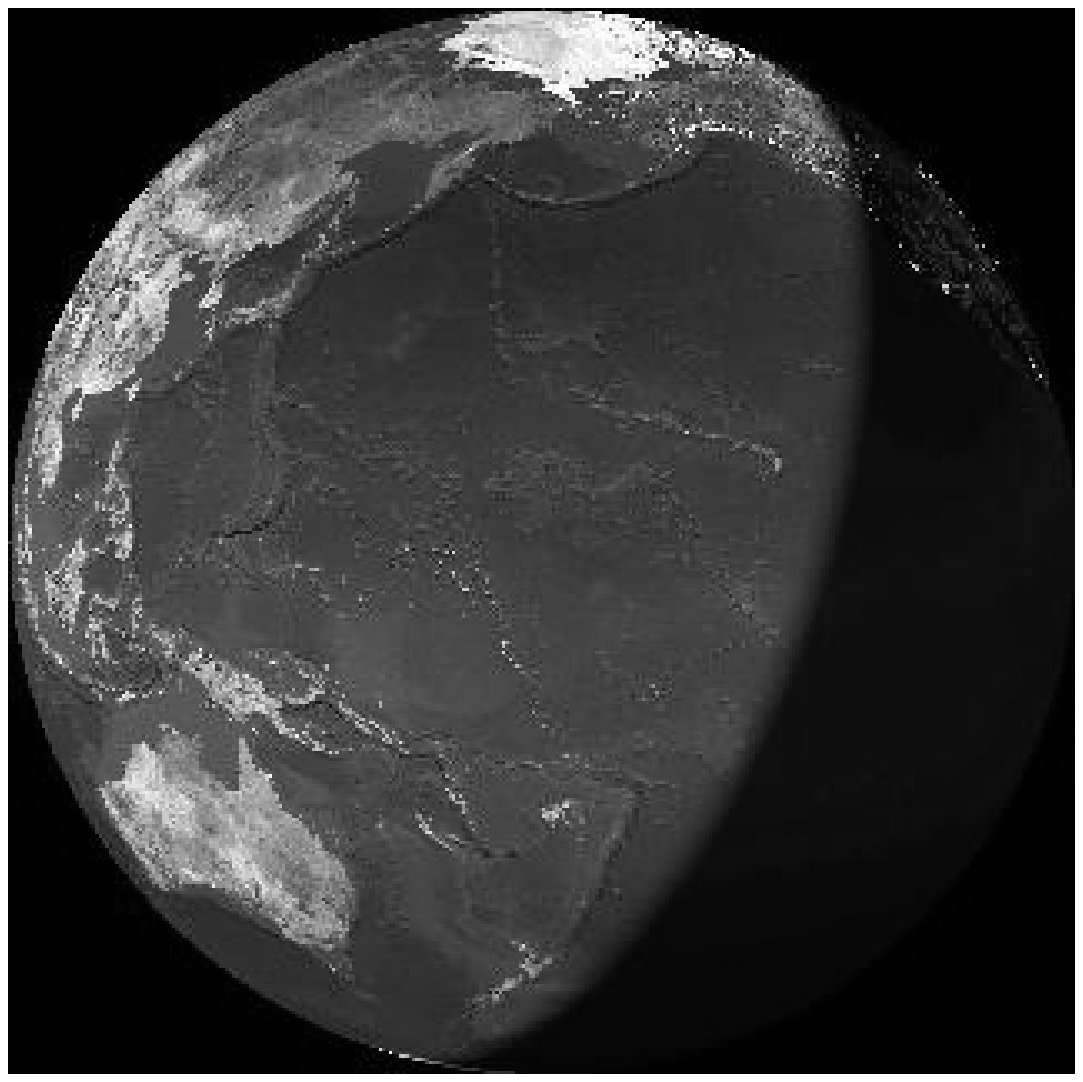}
\caption{A simulated image of the cloud-free, illuminated Earth as seen from the moon 
at the time of the observations reported here, centered on the Pacific Ocean. 
For reference the illuminated appearance of the moon from the Earth would be
approximately the inverse, i.e., the illuminated and dark fractional areas 
would be exchanged.  
}
\end{figure}





\begin{thebibliography}{}

\bibitem[Des Marais et al.(2002)]{desmarais02} 
Des Marais, D., Harwit, M., Jucks, K. W., Kasting, J.,  Lunine, J.,   
Lin, D.,  Seager, S., Schneider, J.,  Traub, W. A., and  Woolf, N. 
2002, Astrobiology, submitted; 
an earlier version is also available as JPL Publication 01-008

\bibitem[Ford et al.(2001)]{ford01}
Ford, E. B., Seager, S., and Turner, E. L.
2001, Nature, 412, 885 

\bibitem[Lane and Irvine(1973)]{lane73}
Lane, A. P. and Irvine, W. M.
1973, \aj, 78, 267

\bibitem[Lunine(1999)]{lunine99}
Lunine, J. I. 1999,
Earth: Evolution of a habitable world,
Cambridge University Press


\bibitem[Traub and Jucks(2002)]{traub02}
Traub, W. A. and Jucks, K. W.
2002, Atmospheres in the Solar System: Comparative Aeronomy,
Edited by M. Mendillo, A. Nagy, and H. J. Waite, 
AGU Geophysical Monograph 130, 000

\end{thebibliography}
\end{document}